\documentclass[aps,prr,twocolumn,groupedaddress,amsmath,epsfig,amssymb,eqsecnum]{revtex4} % nofootinbib
\usepackage{makecell}
\usepackage{bbm}
\usepackage{mathrsfs}
\usepackage{subfigure}
\usepackage{enumerate}
\usepackage{amsfonts}
\usepackage{color}
\usepackage{parskip}
\usepackage{graphicx}% Include figure files
\newcommand{\xing}[1] {{\color{black} #1 }}
\newcommand{\dbar}{d\hspace*{-0.08em}\bar{}\hspace*{0.1em}}
\usepackage{relsize}

\newcommand{\Xv}{\mathbf X}
\newcommand{\Yv}{\mathbf Y}
\newcommand{\Zv}{\mathbf Z}

\newcommand{\xv}{\boldsymbol x}

\newcommand{\zv}{{\mathbf z}}

\newcommand{\qv}{{\boldsymbol q}}
\newcommand{\pv}{{\boldsymbol p}}
\newcommand{\yv}{{\boldsymbol y}}

\newcommand{\be}{\begin{equation}}
\newcommand{\ee}{\end{equation}}
\newcommand{\ba}{\begin{eqnarray}}
\newcommand{\ea}{\end{eqnarray}}

\makeatletter
\newdimen\scalemath@axis
\newcommand*{\scalemath}[3]{%
  #1{%
    \mathpalette{\scalemath@aux{#2}}{#3}%
  }%
}
\newcommand*{\scalemath@aux}[3]{%
  \begingroup
    \everyvbox{}%
    \settoheight\scalemath@axis{$#2\vcenter{}$}%
    \raisebox{\scalemath@axis}{%
      \scalebox{#1}{%
        \raisebox{-\scalemath@axis}{%
          $\m@th#2#3$%
        }%
      }%
    }%
  \endgroup
}
\makeatother 

     \usepackage{boondox-cal}
\begin{document}
	%\preprint{APS/123-QED}
	%Emergent structures
\title{Strong Coupling Thermodynamics and Stochastic Thermodynamics \\ from the Unifying Perspective of Time-Scale Separation}
%\author{Mingnan Ding$^{1}$}
%\author{Zhanchun Tu$^{2}$}
\author{Mingnan Ding$^{1}$}
\author{Zhanchun Tu$^{2}$}
\author{Xiangjun Xing$^{1,3,4}$}
\email{xxing@sjtu.edu.cn}
\address{$^1$Wilczek Quantum Center, School of Physics and Astronomy, Shanghai Jiao Tong University, Shanghai 200240, China \\
$^2$Department of Physics, Beijing Normal University, Beijing 100875, China\\
$^3$T.D. Lee Institute, Shanghai Jiao Tong University, Shanghai 200240, China\\
$^4$Shanghai Research Center for Quantum Sciences, Shanghai 201315, China}
	%\affiliation{}
\date{\today} %freeze this upon submission
	% but any date may be explicitly specifie
	%\pacs{82.70.Dd, 83.80.Hj, 82.45.Gj, 52.25.Kn}
	
	% [CHECK: need to add PACS numbers]
	%\pacs{61.41.+e, 61.43.Er, 62.20.Dc, 64.60.Ak, 64.60.Fr, 64.70.Dv}% PACS, the 
	%61.41.+e 61.43.-j
	%Physics and Astronomy
\begin{abstract}
Assuming time-scale separation, a simple and unified theory of thermodynamics and stochastic thermodynamics is constructed for small classical systems strongly interacting with their environments in a controllable fashion.   The total Hamiltonian is decomposed into a bath part and a system part, the latter being the {\em Hamiltonian of mean force}.   Both the conditional  equilibrium  of bath and the reduced  equilibrium  of the system are described by canonical ensemble theories with respect to their own Hamiltonians.  The bath free energy is independent of the system variables and the control parameter.  Furthermore, the weak coupling theory of stochastic thermodynamics becomes applicable {\em almost verbatim}, even if the interaction and correlation between the system and its environment are strong and varied externally.  We further discuss a simple scenario where the present theory fits better with the common intuition about system entropy and heat.  

% This robustness of equilibrium statistical mechanics and stochastic thermodynamics against strong coupling and large fluctuations is both surprising and 

%Finally, this TSS-based approach also leads to some new insights about the origin of the second law of thermodynamics.  

\end{abstract}

%\vspace{-2mm}

\maketitle 
\section{Introduction }

One of the most significant discoveries of statistical physics in the past few decades is that thermodynamic variables can be defined on the level of dynamic trajectory~\cite{seifert2005entropy,Seifert-review,Jarzynski-review}.  Studies of these fluctuating quantities in non-equilibrium processes have led to significant results such as Fluctuation Theorems~\cite{Seifert-review}, Jarzynski equality~\cite{Jarzynski-review}, as well as a much deeper understanding of the second law of thermodynamics. 

Consider, for example, a small classical system with Hamiltonian $H(\xv, \lambda)$ weakly interacting with its bath, such that the interaction energy and statistical correlation between the system and the bath are negligibly small.  Here $\xv = (\qv, \pv)$ be the canonical variables, and $\lambda$ an external control parameter.  The differential work and heat at trajectory level are defined  respectively as:
\xing{\begin{subequations}
 \ba
\dbar {\mathcal W} &\equiv& 
H(\xv, \lambda + d \lambda) - H(\xv, \lambda)
\equiv d_{\lambda} H(\xv, \lambda), 
\label{dW-def-path} \quad \quad \\ 
\dbar {\mathcal Q}  &\equiv&  H(\xv+d \xv, \lambda) - H(\xv, \lambda)
\equiv d_{\xv} H(\xv, \lambda).  
\label{dQ-def-path} 
\ea
\label{dW-dQ-def-path} 
\end{subequations}
Through out the work, we use the notations $d_{\lambda} H(\xv, \lambda)$ and $d_{\xv} H(\xv, \lambda)$ respectively for differentials of $H(\xv, \lambda) $ due to variations of $\lambda$ and of $\xv$~\footnote{In general, we calculate these differentials up to first order in $dt$, the differential time.  If $\lambda(t), \xv(t)$ are both differentiable,  we have $d_{\lambda} H(\xv, \lambda) = \frac{\partial H}{\partial \lambda} d\lambda, \quad d_{\xv} H(\xv, \lambda) = \frac{\partial H}{\partial \xv} d\xv.  $ If $\xv(t)$ is non-differentiable, which happens if $\xv(t)$ is a typical dynamic trajectory of Brownian motion, then we may need to expand $d_{\xv} H(\xv, \lambda) $ up to the second order in $d\xv$.  In any case, the cross term which is proportional to $d\xv d\lambda$ is always negligible. }.  These notations will greatly simplify the presentation of our theory.  With $H(\xv, \lambda)$ identified as the fluctuating internal energy, the first law at trajectory level then follows directly: $dH = d_{\lambda} H + d_{\xv} H = \dbar {\mathcal W} + \dbar {\mathcal Q} $.   Further using the time-reversal symmetry of Hamiltonian dynamics or Langevin dynamics, one can derive Crooks function theorem, Jarzynski equality, as well as Clausius inequality.  Mathematical expressions for various thermodynamic variables of weak-coupling stochastic thermodynamics are shown in the center column of Table I of Sec. V.   For pedagogical reviews, see e.g. Refs.~\cite{Seifert-review,Jarzynski-review}.  

In recent years,  there have been significant interests in generalizing thermodynamics and stochastic thermodynamics to small systems that are strongly coupled to environment, both classical~\cite{Hanggi-2016-strong-coupling,Jarzynski-2017-strong-coupling,Seifert-2016-strong-coupling,Strasberg-2016-strong-coupling,Gelin-2009,Hanggi-2020-Colloquium,Strong-Miguel-2020,Strong-Aurell-2018,Strong-Anders-2017,Esposito-strong-coupling-2017,Strasberg-2020-HMF,Campisi-2009-HMF}, and quantum~\cite{Hanggi-2016-strong-coupling,Hanggi-2020-Colloquium,Strong-Huang-20120,Strong-Hsiang-20121,Strong-Carrega-2016,Rivas-2020-strong-coupling,Perarnau-Llobet-2018-strong-coupling,Strasberg-2019-strong-coupling,Pucci-2013-poised,Esposito-2015-heat,Ptaszynski-2019,ELB-2010,Abe-2003-strong-coupling,Miguel-Rubi-2020}.  Strong interactions between system and its environment cause ambiguities the definitions of system thermodynamic quantities~\cite{Hanggi-2016-strong-coupling,Hanggi-2020-Colloquium}.  If the system size is large, and the interactions are short-ranged, the correlations between system and bath are confined to the interfacial regions, and hence do not influence the bulk properties of the system.  This is indeed the reason why classical thermodynamics and statistical mechanics are so successful in describing the equilibrium properties of macroscopic systems, even if these systems may be strongly interacting with environment near the interfaces.   Small systems however have no ``bulk'', and their thermodynamic properties may be overhelmly dominated by their interactions and correlations with environment.  Whether one should relegate the interaction energy to the system or to the bath? Whether one should treat the mutual information between system and hath variables as part of system entropy or bath entropy?  There seems no general principle in favor of any particular answer.  For critical and insightful discussions of these fundamental issues, see the recent articles by Jarzynski~\cite{Jarzynski-2017-strong-coupling}, and by Talkner and H\"anggi~\cite{Hanggi-2020-Colloquium}.  

Numerous versions~\cite{Seifert-2016-strong-coupling,Hanggi-2016-strong-coupling,Gelin-2009,Roux-solvent-1999,Jarzynski-2017-strong-coupling}  of strong coupling thermodynamic theories have been proposed in recent years.  Probably the most influential theory was developed by Seifert~\cite{Seifert-2016-strong-coupling}, and critically evaluated by Talkner and H\"anggi~\cite{Hanggi-2016-strong-coupling,Hanggi-2020-Colloquium}.  In this theory, one uses the {\em Hamiltonian of mean force} (HMF) $H_{\Xv}$~\cite{Jarzynski-2004-HMF,Campisi-2009-HMF,Kirkwood-1935} to construct the equilibrium free energy $F = - T \log \int e^{- \beta H_{\Xv}}$,  and then {\em defines} equilibrium system energy and entropy via $  E = {\partial \beta F}/{\partial \beta},   S = - \beta^2 {\partial F}/{\partial \beta} $. Whilst these relations exactly hold in equilibrium thermodynamics, they must be deemed as {\em definitions} of energy and entropy in Seifert's theory of strong coupling thermodynamics.  Interestingly, these definitions correspond to the particular decomposition of total thermodynamic variables $A_{\rm tot} = A_{\rm sys} + A_{\rm bath}$, where $A_{\rm bath}$ is the thermodynamic variable of the {\em bare bath}, with the interaction between the system and bath switched off.  Hence it can be said that Seifert allocates the entire interaction energy to the system.  These definitions of energy and entropy are further bootstrapped to non-equilibrium situations~\cite{Seifert-2016-strong-coupling}, and fluctuation theorems and Clausius inequality are subsequently established.  The resulting formulas (the right column of Table I) in strongly coupled regimes are markedly more complicated than those in weak coupling theory (the central column).  These differences however disappear as the interaction Hamiltonian vanishes, and the HMF reduces to the bare system Hamiltonian.   

Strasberg and Esposito~\cite{Esposito-strong-coupling-2017} recently studied the strong coupling problem  from the viewpoint of time-scale separation (TSS).  They consider a system involving both slow and fast variables.  By assuming fast variables in conditional equilibrium, they show that Seifert's theory can be derived by averaging out the fast variables.  Furthermore, they proposed a definition of total entropy production in terms of relative entropy, which is a variation of entropy production defined Ref.~\cite{Esposito-2010}, and show that it is equivalent to the entropy production in Seifert's theory.  The conditional equilibrium of bath also allows one to prove the positivity of instantaneous rate of total entropy production, rather than the positivity of total entropy production of an entire process.  The importance of TSS has long been known. It was invoked heuristically to justify adiabatic approximation~\cite{Born-Fock-1928-adiabatic,Born-Oppenheimer-1927}, Markov modeling~\cite{Gardiner-book}, or dimensional reduction of dynamic theories~\cite{Michaelis-Menten-1913,Pavliotis-Stuart-book-2008}.  

Jarzynski~\cite{Jarzynski-2017-strong-coupling} developed a more comprehensive (and hence more complex) theory for strong coupling thermodynamics, and systematically discussed the definitions of internal energy, entropy, volume, pressure, enthalpy, and Gibbs free energy.  The formalism was established around the concept of volume, whose definition is somewhat arbitrary.   All other thermodynamic variables are uniquely fixed by thermodynamic consistency once the system volume is (arbitrarily) defined. Jarzynski further showed that Seifert's theory is a special case of his framework, i.e., the ``partial molar representation''.  He discussed in great detail the ``bare representation'', where the system enthalpy coincides with HMF.  The total entropy production is however the same in both representations.   Jarzynski made analogy between the arbitrariness in the definition of thermodynamic variables in the strong coupling regime and the gauge degree of freedom in electromagnetism, which was criticized by Talkner and H\"anggi~\cite{Hanggi-2020-Colloquium}. 

The main purpose of the present work is to show that, with TSS and the ensuing conditional equilibrium of bath variables, a much simpler thermodynamic theory can be developed for strongly coupled small classical systems.  More specifically, we will show that by identifying the {\em Hamiltonian of Mean Force} (HMF) as the system Hamiltonian, and relegating the remaining part of the total Hamiltonian to the bath, both the equilibrium ensemble theory and the weak coupling theory of stochastic thermodynamics remain applicable, {\em almost verbatim}, in the strong coupling regime.   Work and heat, entropy, and energy all retain the same definitions and the same physical meanings as in the weak coupling theory, as long as the bath entropy understood as conditioned on the system state.   Fluctuation Theorems, Jarzynski equality, and Clausius inequality can all be proved using nonlinear Langevin dynamics~\cite{covariant-Thermodynamics-2021}, whose validity relies on TSS but not on strength of coupling.  Using the conditional equilibrium nature of bath, it can be rigorously demonstrated that $dS - \beta \dbar Q$ equals to the entropy change of the universe, which establishes the meaning of Clausius inequality as increasing total entropy.  Finally, we will also show that our theory, though significantly simpler, are consistent with all previous theories, in the sense that the total entropy productions in all theories are mathematically equivalent.  Summarizing, we achieve a natural unification of thermodynamics and stochastic thermodynamics at both weak and strong coupling regimes.

A logical consequence of TSS is that the dynamic evolution of slow variables can modeled as Markov process, such as Langevin dynamics with white noises.  In the strongly coupled regime, the noises are however generically multiplicative.  In a complementary paper~\cite{covariant-Thermodynamics-2021}, two of us develop a theory of stochastic thermodynamics using nonlinear Ito-Langevin dynamics, establish its covariance property, and derive Crooks Fluctuation Theorem, Jarzynsk equality, as well as Clausius inequality.  The definitions of thermodynamic quantities are identical in these two works, if we take $g_{ij} = \delta_{ij}$ in Ref.~\cite{covariant-Thermodynamics-2021}. (The theory in Ref.~\cite{covariant-Thermodynamics-2021} was developed for Langevin dynamics on arbitrary Riemannian manifold with invariant volume measure $\sqrt{g}\, d^dx$, whereas in the present work, we consider Hamiltonian systems with Liouville measure $\prod_i dp_idq_i$.  )  Combination of these two works provides a covariant theory of thermodynamics and stochastic thermodynamics for systems strongly interacting with a single heat bath, with TSS as the only assumption.  
}
%\vspace{1mm}

The remaining of this work is organized as follows.  In Sec.~\ref{sec:EQ}, we introduce our decomposition of the total Hamiltonian, and discuss the equilibrium thermodynamic properties of strongly coupled systems.  In Sec.~\ref{sec:NEQ}, we discuss the non-equilibrium thermodynamic properties of the system.   Work and heat retain the same definitions and same physical meanings as in the weak coupling theory, i.e.,  the energy changes of the combined system and of the bath respectively.  In Sec.~\ref{sec:heat}, we discuss the connection between heat and entropy change of the bath, conditioned on the slow variables.  In Sec.~\ref{sec:comparison}, we compare our theory with previous theories by Seifert, by H\"anggi and Talkner, by Jarzynski, and by Strasberg and Esposito, and show that they are all equivalent.  We will also discuss a simple scenario where the present theory fits better with the common intuition about system entropy and heat.  In Sec.~\ref{sec:conclusion} we draw concluding remarks.

\vspace{-2mm}
\section{Equilibrium  Theory }
\label{sec:EQ}
In this section, we shall demonstrate that by identifying HMF as the system Hamiltonian, and the remaining of the total Hamiltonian as the bath Hamiltonian, canonical ensemble theory can be straightforwardly adapted to describe the equilibrium properties of systems that are strongly coupled to their baths.  There is  also a related decomposition of total thermodynamic quantities into system parts and bath parts.  The bath free energy turns out to be the same as that of a bare bath, and is independent of the state of slow variables or of the external control parameter.   
%\vspace{-4mm}

\xing{
\subsection{Decomposition of total Hamiltonian}
\label{sec:H-decomp}
 We shall use $\Xv, \Yv$ to denote fast and slow variables, and $\xv, \yv$ their values.  We shall also call $\Xv$ {\em the system} and $\Yv$ {\em the bath}.  Let the total Hamiltonian be 
\ba
H_{\Xv\Yv} (\xv, \yv; \lambda)  &=& H_{\Xv}^0(\xv; \lambda)
 + H_{\Yv}(\yv)  + H_I^0(\xv, \yv; \lambda),
\nonumber\\
 \label{H_tot-decomp-1}
\ea
where $H_{\Xv}^0(\xv; \lambda)$ and $H_{\Yv}(\yv)$ are the {\em bare system Hamiltonian} and {\em bare bath Hamiltonian}, whereas $H_I^0(\xv, \yv; \lambda)$ is the {\em bare interaction}.  Note that every term in the RHS is independent of temperature, and the { bare bath Hamiltonian} $H_{\Yv}(\yv)$ is independent of $\lambda$.  Our starting point Eq.~(\ref{H_tot-decomp-1}) is more general than those in Ref.~\cite{Seifert-2016-strong-coupling,Hanggi-2016-strong-coupling,Jarzynski-2017-strong-coupling}, where the  bare interaction $H_I^0(\xv, \yv; \lambda)$ is assumed to be independent of $\lambda$.

Throughout this work, we shall assume that $\Xv\Yv$ is weakly interacting with a much larger super-bath whose dynamics is even faster than $\Yv$.  We will call $\Yv\Zv$ the {\em environment}, and $\Xv\Yv\Zv$ the {\em universe}.
We shall use $\int_{\yv} \equiv \int d^N \! y$ to denote integration over $\yv$, and similar notation for integration over $\xv$ and $\zv$.  These notations are especially useful when we dealing with integrals of differential forms.}  \xing{Let $T = 1/\beta$ be the temperature, which is assumed to be fixed throughout the work.  We shall set the Boltzmann constant $k_B = 1$, and hence all entropies are dimensionless. 

 We shall define {\em system Hamiltonian} $H_{\Xv} (\xv; \lambda, \beta)$ and {\em interaction Hamiltonian} $H_{I}(\xv, \yv; \lambda, \beta)$ as
\ba
H_{\Xv} (\xv; \lambda, \beta)  &=&  H_{\Xv}^0(\xv) 
-  T \log \frac { \int_{\yv} e^{-\beta (H_{\Yv} + H_{I}^0 )} }
{ \int_{\yv} e^{-\beta H_{\Yv}}}
\nonumber\\
&=&-  T \log \frac { \int_{\yv} e^{-\beta H_{\Xv\Yv}  } }
{ \int_{\yv} e^{-\beta H_{\Yv}}},
\label{xxx-app-1} \\
H_{I}(\xv, \yv; \lambda, \beta) &=&
 H_{I}^0(\xv, \yv;\lambda) 
 +  T \log \frac{ \int_{\yv} e^{-\beta H_{\Yv}}}
{ \int_{\yv} e^{-\beta (H_{\Yv} + H_{I}^0 )} },
\nonumber\\
\label{xxx-app-2}
\ea
both of which depend  on $\beta$ and $\lambda$.   Note that $H_{\Xv} (\xv; \lambda, \beta) $ is precisely the {\em Hamiltonian of mean force} (HMF) defined and used in previous works~\cite{Jarzynski-2004-HMF,Campisi-2009-HMF,Seifert-2016-strong-coupling,Hanggi-2016-strong-coupling}~\footnote{Recall that we are dealing with classical systems, where there is no problem of non-commutativity.  For quantum systems, Eq.~(\ref{xxx-app-1}) is equals to Hamiltonian of mean force only if $H_{\Xv}^0$ commute with $H_{\Yv} + H_I^0$. }.  

We now obtain  a new decomposition of $H_{\Xv\Yv}$:
\begin{subequations}
\ba
H_{\Xv\Yv} (\xv, \yv; \lambda) &=& H_{\Xv}(\xv; \lambda, \beta)
 + H_{\Yv}(\yv)  + H_I(\xv,  \yv; \lambda, \beta).  
 \nonumber\\
\label{H_tot-decomp-2} 
\label{H_tot-decomp-0}
\ea
Note that even though both $H_{\Xv}$ and $H_{I}$ depend on $\beta$, the total Hamiltonian in LHS of Eq.~(\ref{H_tot-decomp-0}) is independent of $\beta$.   We further define {\em bath Hamiltonian} as
\be
 H_{\rm Bath}(\yv;\xv,\lambda, \beta) \equiv H_{\Yv}(\yv)  
 + H_I(\xv,  \yv; \lambda, \beta),
\label{H_Bath-def}
\ee  
and rewrite Eq.~(\ref{H_tot-decomp-2}) as
\ba
H_{\Xv\Yv} (\xv, \yv; \lambda)  &=& 
H_{\Xv}(\xv; \lambda, \beta) 
+ H_{\rm Bath}(\yv;\xv,\lambda, \beta). 
\nonumber\\
\label{H_tot-decomp}
\ea
\label{H_bath-def}
\end{subequations}
We also define the {\em bath partition function} as
\ba
Z_{\Yv} (\xv, \lambda, \beta)  &=& \int_{\yv} \,  
e^{-\beta H_{\rm Bath}(\yv;\xv,\lambda, \beta) },
\label{Z_Y-def-0}
\ea
which is conditioned on $\Xv = \xv$, and generally also depends on both $\xv$ and $\lambda$.     Using Eqs.~(\ref{H_Bath-def}) and (\ref{xxx-app-2}), we easily see that
\ba
Z_{\Yv} (\xv, \lambda, \beta)  &=&  \int_{\yv} e^{-\beta (H_{\Yv} + H_{I}^0 )} 
 \frac{ \int_{\yv} e^{-\beta H_{\Yv} } }
{ \int_{\yv} e^{-\beta (H_{\Yv} + H_{I}^0 )}  }
\nonumber\\
&=& \int_{\yv} e^{-\beta H_{\Yv} }  \equiv Z_{\Yv}^0(\beta),  
 \label{Z_Y-def-2}
\ea
where $ Z_{\Yv}^0(\beta) $ is the partition function of the bare bath, with the interaction Hamiltonian between $\Xv$ and $\Yv$ completely switched off.  

Hence the bath partition function $Z_{\Yv} (\xv, \lambda, \beta) $ as defined by Eq.~(\ref{Z_Y-def-0}) is independent of $\xv$ and $\lambda$:
%\begin{subequations}
\ba
\frac{\partial Z_{\Yv}   (\xv, \lambda, \beta) }{\partial \xv}   = 
\frac{\partial Z_{\Yv}   (\xv, \lambda, \beta)  }{\partial \lambda}  = 0,
%\frac{\partial}{\partial \xv} \int_{\!\! \yv} \,  
%e^{-\beta H_{\Yv}(\yv) - \beta H_I(\xv, \yv)} = 0.  
\label{H_I-constraint-1}
\label{H_I-constraint-12}
\ea
%\end{subequations}
and we shall from now on simply write it as  $Z_{\Yv}  (\beta)$.  Equation (\ref{H_I-constraint-12}) will play a very significant role in our theory.  
}

\subsection{Conditional  equilibrium  of bath} 
\label{Sec:cond-CE}

In an intermediate time-scale, the fast variables equilibrate whereas the slow variables barely change.  Hence $\Yv$ achieves  equilibrium {\em conditioned on $\Xv = \xv$}, described by the conditional Gibbs-Boltzmann distribution:
\begin{subequations}
\ba
p^{\rm EQ}_{\Yv|\Xv}(\yv|\xv) &=& \frac{1}{Z_{\Yv} (\beta) } 
e^{-\beta H_{\rm Bath}(\yv;\xv,\lambda, \beta) }, 
\label{p_Y-EQ-1}
\ea
with $Z_{\Yv} (\beta)$ defined in Eq.~(\ref{Z_Y-def-0}).  We further define  the conditional free energy of the bath:
\ba
F_{ \Yv} (\beta) &\equiv& - T \log Z_{\Yv} (\beta) 
= -T \log Z_{\Yv}^0(\beta) = F_{ \Yv}^0 (\beta), 
\nonumber\\
% &=& - T \log \int_{\yv} \, e^{-\beta H_{\rm Bath}(\yv;\xv,\lambda, \beta) },
\label{Z_Y-def}
\ea
 \label{F-Z_Y}
\end{subequations}
where $F_{ \Yv}^0 (\beta)$ is the free energy of the bare bath.  Equations (\ref{F-Z_Y}) define a {\em conditional canonical ensemble}, which describes the  equilibrium physics of the fast variables in the intermediate time scales, during which the slow variables are change very little.  In this ensemble theory, $\xv$ serves as a parameter, just like $\lambda$ and $\beta$.

The  internal energy and entropy of bath in the conditional  equilibrium state are defined in a standard way:
\begin{subequations}
\ba
E_{\Yv}(\xv) &=& %\langle ( H_{\Yv} + H_{I} ) \rangle 
   \int_{\yv}\, p^{\rm EQ}_{\Yv|\Xv}(\yv|\xv) 
   H_{\rm Bath}(\yv;\xv,\lambda, \beta) , 
    \label{E_Y-X-cond}\\
S_{\Yv|\Xv = \xv} &=& - \int_{\yv}\, 
 p^{\rm EQ}_{\Yv|\Xv}(\yv|\xv) 
\log p^{\rm EQ}_{\Yv|\Xv}(\yv|\xv),    
\label{S_Y-X-cond}
\ea
which are related to the free energy $F_{ \Yv} (\beta) $ via
\ba
F_{ \Yv} (\beta) &=& E_{\Yv}(\xv)  - T S_{\Yv|\Xv = \xv} . 
\label{F_Y-E-S-x}
\ea
\label{TMV-Y-1}
\end{subequations}
\xing{ $S_{\Yv|\Xv = \xv} $ is known in information theory~\cite{Cover-Thomas} as {\em the conditional Shannon entropy of $\Yv$ given $\Xv = \xv$}. } Note that even though $F_{ \Yv} (\beta) $ does not depend on $\xv$ and $\lambda$, both $ E_{\Yv}(\xv)$ and $S_{\Yv|\Xv = \xv}$ do.  
% Experimentally measurable quantities can always be expressed in terms of free energy $F_{ \Yv} (\beta)$.   

Even though the  free energy of the bath conditioned on the system state equals to that of the bare bath, there are important differences between other thermodynamic quantities of the bath and the bare bath. For example, the internal energy and entropy of the bare bath are given respectively by
\begin{subequations}
\ba
E^0_{\Yv} &=& \int_{\yv}\, 
\frac{e^{-\beta H_{\Yv}}}{Z_{\Yv}^0(\beta)}    H_{\Yv}(\yv) , 
    \label{E_Y-bare}\\
S_{\Yv}^0 &=& - \int_{\yv}\, 
\frac{e^{-\beta H_{\Yv}}}{Z_{\Yv}^0(\beta)}  
\log \frac{e^{-\beta H_{\Yv}}}{Z_{\Yv}^0(\beta)}  ,    
\label{S_Y-bare}
\ea
\label{TMV-Y-0}
 \end{subequations}
which are manifestly different from Eqs.~(\ref{E_Y-X-cond}), (\ref{S_Y-X-cond}).  

\subsection {Joint  equilibrium  of system and bath} 
\label{sec:joint-EQ}
In long time-scales, $\Xv\Yv$ achieve a joint  equilibrium, which is described by the joint Gibbs-Boltzmann distribution: 
\ba
p^{\rm EQ}_{\Xv\Yv}(\xv, \yv) &=& 
\frac{e^{-\beta H_{\Xv\Yv}(\xv, \yv; \lambda) }}
{Z_{\Xv\Yv}(\beta, \lambda)},
\label{p-EQ-XY}
\ea
where $Z_{\Xv\Yv}(\beta, \lambda)$ is the canonical joint partition function:
\begin{subequations}
\ba
Z_{\Xv\Yv}(\beta, \lambda)
 &=& \int_{ \xv\yv}\!\! e^{-\beta H_{\Xv\Yv}}
\nonumber\\
 &=& \int_{ \xv\yv}\!\! e^{-\beta H_{\Xv} -  \beta H_{\rm bath} }.  
%\!\!  \! \int_{\yv} \!\! e^{-\beta H_{\rm Bath}}, 
%=  \int_{\yv}\,e^{-\beta H_{\Xv}} Z_Y(\beta). \quad
 \label{Z_XY-prod-0}
\ea
From this we can obtain various thermodynamic quantities for this {\em joint canonical ensemble} in a standard way:
\ba
F_{\Xv \Yv}(\beta, \lambda)  &=& 
-T \log Z_{\Xv\Yv}(\beta, \lambda), 
% - T \log  \int_{\!\!\xv\yv} \!\! e^{-\beta H_{\Xv\Yv}} ,
\label{F-Z_XY} \\
E_{\Xv\Yv}(\beta, \lambda) &=&% \langle H_{\Xv} \rangle  = 
\int_{\xv,\yv} p^{\rm EQ}_{\Xv\Yv}(\xv, \yv) 
H_{\Xv\Yv}(\xv, \yv; \lambda) , 
\label{E_XY-1}\\
S_{\Xv\Yv}(\beta, \lambda) &=&  - \int_{\xv,\yv}\!\! 
p_{\Xv\Yv}^{\rm EQ} (\xv, \yv)
\log p_{\Xv\Yv}^{\rm EQ}(\xv, \yv), \quad \quad
\label{S_XY-canonical-0}\\
F_{\Xv\Yv}(\beta, \lambda) &=& E_{\Xv\Yv} - T S_{\Xv\Yv}. 
\ea
\label{TMV-XY-1}
\end{subequations}
The joint canonical ensemble describes the equilibrium statistical properties of both slow and fast variables.  

%\vspace{-3mm}
\subsection{ Reduced  equilibrium of System} 

But we may also study the equilibrium distribution of slow variables alone. This {\em reduced canonical distribution} can be obtained from Eq.~(\ref{p-EQ-XY}) by integrating out the fast variables: 
\ba
p^{\rm EQ}_{\Xv}(\xv) &=& \int_{\yv}\, p^{\rm EQ}_{\Xv\Yv}(\xv, \yv)
\nonumber\\
&=& \frac{1}{Z_{\Xv\Yv}(\beta, \lambda) } \int_{\yv} e^{-\beta H_{\Xv} -  \beta H_{\rm bath} }  
\nonumber\\
&=& \frac{Z_{\Yv}(\beta) }{Z_{\Xv\Yv}(\beta, \lambda) }
e^{-\beta H_{\Xv}},
\label{p_X-HMF-1} 
\ea
where used was Eq.~(\ref{Z_Y-def-0}) and the fact that $Z_{\Yv}(\beta)$ is independent of $\xv$.  Hence the equilibrium distribution of $\Xv$ is canonical with respect to the system Hamiltonian $H_{\Xv}(\xv; \lambda)$.  This is, of course, well known, since we have defined $H_{\Xv}(\xv; \lambda)$ as the HMF.

It is then convenient to define the partition function of slow variables:
\ba
Z_{\Xv}(\beta, \lambda) 
&\equiv& \int_{\xv} e^{-\beta H_{\Xv}(\xv; \lambda, \beta)},
%= \frac{Z_{\Xv\Yv}(\beta, \lambda)}{Z_{\Yv}(\beta)} .  
\label{Z_X-def}
\ea
so that Eq.~(\ref{p_X-HMF-1}) assumes the standard canonical form:
\ba
p^{\rm EQ}_{\Xv}(\xv) &=&  \frac{1 }{Z_{\Xv}(\beta, \lambda) }
e^{-\beta H_{\Xv}}. 
\label{p_X-HMF-2} 
\ea
Integration of Eq.~(\ref{p_X-HMF-1}) then yields
\begin{subequations}
\ba
Z_{\Xv\Yv}(\beta, \lambda)  &=&
 Z_{\Xv}(\beta, \lambda)  Z_{\Yv}(\beta). 
\label{Z-factorization-1}
%&=& Z_{\Xv}(\beta, \lambda)  Z^0_{\Yv}(\beta). 
%\label{Z-factorization-2}
\ea 
\end{subequations}

The above results prompt us to construct a {\em reduced canonical ensemble} theory for the system, with free energy, internal energy, and entropy given by
\begin{subequations}
%\vspace{3mm}
\label{reduced-X-thermodynamics}
\ba
F_{\Xv } &=& -T \log Z_{\Xv}(\beta, \lambda), 
\label{F-Z_X}\\
 E_{\Xv} &=&% \langle H_{\Xv} \rangle  = 
\int_{\xv} \, p^{\rm EQ}_{\Xv}(\xv) 
 H_{\Xv}(\xv; \lambda), 
\label{E_X-1}\\
S_{\Xv} &=& - \int_{\xv} \,p_{\Xv}^{\rm EQ}(\xv) 
\log p_{\Xv}^{\rm EQ}(\xv),
\label{S_X-canonical-0}\\ 
F_{\Xv} &=& E_{\Xv} - T S_{\Xv}. 
\label{F-E-S-X}
\ea
\end{subequations}

These definitions of system energy and entropy  are manifestly different from the strong coupling theory in Refs.~\cite{Hanggi-2016-strong-coupling,Seifert-2016-strong-coupling,Hanggi-2020-Colloquium}, even though the free energy is the same in two theories.  

\vspace{-4mm}
\subsection{Decomposition of Thermodynamic Variables}
\label{sec:decomp-TMV}

Comparing Eqs.~(\ref{reduced-X-thermodynamics}) with Eqs.~(\ref{TMV-Y-1}) and (\ref{TMV-XY-1}), we find the following decomposition of total thermodynamic quantities into system parts and bath parts:
\begin{subequations}
\ba
F_{\Xv\Yv}(\beta, \lambda) &=& F_{\Xv}(\beta, \lambda) + F_{\Yv}(\beta), 
\label{F_XY-canonical}\\
E_{\Xv\Yv} &=&  E_{\Xv} + \langle E_{\Yv}(\xv) \rangle_{\Xv}, \\
S_{\Xv\Yv}  &=&   S_{\Xv} + S_{\Yv|\Xv},  
\label{S_XY-canonical}
\ea
\label{TMV-canonical}
\label{Additivity-1}
\end{subequations}
where $ \langle E_{\Yv}(\xv) \rangle_{\Xv}$ and $S_{\Yv|\Xv}$ are respectively averages of $E_{\Yv}(\xv)$ and  $S_{\Yv|\Xv = \xv}$ over fluctuations of $\Xv$:
\begin{subequations}
\ba
 \langle E_{\Yv}(\xv) \rangle_{\Xv} 
 &=& \int_{\xv} \, p^{\rm EQ}_{\Xv}(\xv) E_{\Yv}(\xv), \\
 S_{\Yv|\Xv}  = 
 \langle S_{\Yv|\Xv= \xv}  \rangle_{\Xv}
 &=&  \int_{\xv} \, p^{\rm EQ}_{\Xv}(\xv) S_{\Yv|\Xv= \xv} . 
  \quad\quad
   \label{S_Y_conditional-1}
\ea
\end{subequations}
\xing{$ S_{\Yv|\Xv} $ is called the {\em conditional Shannon entropy of $\Yv$ given $\Xv$} in information theory~\cite{Cover-Thomas}.  Note the subtle differences between the names for $ S_{\Yv|\Xv} $ and for $ S_{\Yv|\Xv = \xv} $. }

\vspace{-4mm}
There are numerous pleasant features of the equilibrium thermodynamic theory developed here: 
Firstly all  equilibrium distributions are Gibbs-Boltzmann with respect to the corresponding Hamiltonian. Secondly, all entropies are Gibbs-Shannon entropy with respect to the corresponding pdfs.  As a consequence, the formulas in Eqs.~(\ref{F-Z_Y}), (\ref{TMV-Y-1}), (\ref{TMV-XY-1}), and (\ref{reduced-X-thermodynamics}) are all the same as those in  canonical ensemble theory.  These feature are remarkable, since they indicate that standard canonical ensemble theory are applicable both to the system and to the bath, regardless of the strong interaction and correlation between them.   Thirdly, Eq.~(\ref{Z_Y-def}) says that the bath free energy $F_{\Yv}(\beta)$ equals to the bare bath free energy $F_{\Yv}^0(\beta)$, and is independent of $\lambda$ and $\xv$.  This feature leads to substantial conceptual simplification since we are only interested in the physics of slow variables.  Consider, for example we immerse a DNA chain into an aqueous solvent, or stretch it in the solvent, or tune the interaction between a nano-engine and its environment.  There is no need to worry about the change of bath free energy, because it stays constant by construction.  

All these convenient features follow from the particular decomposition of total Hamiltonian Eqs.~(\ref{H_bath-def}).   There are however some subtleties resulting from the temperature dependence of $H_{\Xv}$, which will be discussed in Sec.~\ref{sec:comparison}.  We shall also give a detailed comparison between our  theory  and the previous theories by Seifert, H\"anggi and Talkner, and by Jarzynski in Sec.~\ref{sec:comparison}.  

%This feature has not been noticed by previous theories.  
 
\section{Non-equilibrium Theory}
\label{sec:NEQ}
In this section, we shall show that with the HMF $H_{\Xv}$ identified as the fluctuating internal energy, the weak coupling theory of stochastic thermodynamics becomes applicable in the strong coupling regime.  

%\vspace{-2mm}
\subsection{Definitions of energy and entropy}
\label{sec:IIIA}
The mission of stochastic  thermodynamics starts with definitions of system thermodynamic variables in general non-equilibrium situations.  We define the fluctuating internal energy of the system as $H_{\Xv}(\xv; \lambda, \beta)$, the HMF.  The non-equilibrium internal energy is then defined as the ensemble average of $H_{\Xv}$:
\be
E_{\Xv}[p_{\Xv}]  \equiv  - \int_{\xv}  p_{\Xv} H_{\Xv}.  
\label{E-ensemble-def}
\ee  
Throughout this work we use $A[p_{\Xv}]$ to denote a non-equilibrium thermodynamic variable, to distinguish it from the equilibrium version $A$.  We also define the system entropy as the Gibbs-Shannon entropy: 
\be
S_{ \Xv} [p_{\Xv}]  \equiv -  \int_{\xv} p_{\Xv}(\xv, t) \log p_{\Xv}(\xv, t).  
\ee  
We shall not need to define stochastic entropy in this work.  The non-equilibrium free energy of the system is also defined in the standard way:
\ba
F_{\Xv}[p_{\Xv}] &=& E_{\Xv}[p_{\Xv}] - T\, S_{ \Xv} [p_{\Xv}] 
\nonumber\\
&=& \int_{\xv} p_{\Xv} \left( 
H_\Xv + T \log p_\Xv \right),
\label{F_X-def} 
\ea
which turns out to be the same as the free energy defined in several previous theories~\cite{Hanggi-2016-strong-coupling,Seifert-2016-strong-coupling,Hanggi-2020-Colloquium,Esposito-strong-coupling-2017}.  

Note that these definitions of non-equilibrium entropy, energy, and free energy are identical to those in weak coupling theory, with $H_{\Xv}$ understood as the system Hamiltonian.  For equilibrium state $p_{\Xv} = p_{\Xv}^{\rm EQ}$, these thermodynamic variables reduce to their equilibrium counterparts, Eqs.~(\ref{E_X-1}), (\ref{S_X-canonical-0}), and (\ref{F-Z_X}) respectively.

\subsection{Work and heat at trajectory level} 
Let us now discuss differential work and heat at trajectory level of system variables. 

The Hamiltonian of the {\em universe}, including system, bath and super-bath, is given by 
\ba
{ H_{\Xv\Yv\Zv}} &=& { H_{\Xv\Yv}} + H_{\Zv} 
\nonumber\\
&=& H_{\Xv} + H_{\rm Bath} + H_{\Zv}, 
\label{H_XYZ-def}
 \ea
with $ { H_{\Xv\Yv}} $ given by Eqs.~(\ref{H_bath-def}).  We assume that the interaction between $\Xv\Yv$ and $\Zv$ is negligibly small but nonetheless is strong enough to drive thermal equilibration between $\Xv\Yv$ and $\Zv$.   

We consider a microscopic process with infinitesimal duration $dt$, where $\xv, \yv,\zv$ and $\lambda$ change by $d\xv, d\yv, d\zv$ and $d\lambda$.   Whereas $d \lambda$ is externally controlled, $d\xv, d\yv, d\zv$ are determined by  evolution of Hamiltonian dynamics.  \xing{As is generally adopted in stochastic thermodynamics, work is defined as the change of total energy of the universe: 
\ba
&& \dbar \tilde{\mathcal W} = d { H_{\Xv\Yv\Zv}}.  
\label{dE-tot-lambda-0}
%\nonumber \\
%&=& H_{\Xv\Yv\Zv} (\xv + d\xv, \yv + d \yv, \zv + d\zv; \lambda + d \lambda)
%-  H_{\Xv\Yv\Zv} (\xv , \yv , \zv ; \lambda )
%  \label{dE-tot-lambda-0} \\
% &=& d_{\xv}{ H_{\Xv\Yv\Zv}} 
% + d_{\yv}{ H_{\Xv\Yv\Zv}} 
%+ d_{\zv}{ H_{\Xv\Yv\Zv}} 
%+ d_{\lambda}{ H_{\Xv\Yv\Zv}}, 
%\nonumber
\ea
For now we shall assume that $\xv, \yv, \zv, \lambda$ are all smooth functions of $t$~\footnote{This is reasonable as long as we do not take the fast time scale to zero. }.  We can then expand Eq.~(\ref{dE-tot-lambda-0}) in terms of $d\xv, \cdots, d \lambda$ up to the first order.  The coefficients are just the partial derivatives of $ H_{\Xv\Yv\Zv}$ with respect to  $d\xv, \cdots, d \lambda$.  Now note that the universe $\Xv\Yv\Zv$ is thermally closed.  Hence  if $\lambda$ is fixed, $H_{\Xv\Yv\Zv}$ must be conserved.  In another word, Eq.~(\ref{dE-tot-lambda-0}) can change only due to $\lambda$:
\ba 
\dbar \tilde{\mathcal W}  &=& 
\frac{\partial H_{\Xv\Yv\Zv}}{\partial \lambda}d \lambda
\equiv  d_{\lambda}  { H_{\Xv\Yv\Zv}} . 
 \label{dE-tot-lambda-1}
\ea
  Further using Eqs.~(\ref{H_XYZ-def}) and (\ref{H_tot-decomp-2}) we can rewrite the preceding equation as 
\ba
\dbar \tilde{\mathcal W}   = d_{\lambda} \left( { H_{\Xv\Yv}}  + H_{\Zv} \right)
=   d_{\lambda}{ H_{\Xv}} + d_{\lambda}{ H_{I}},
\label{dW-def-1}
\ea
where in the last equality we have used the fact that both $H_{\Yv}$ and $H_{\Zv}$ are independent of $\lambda$.  Hence the microscopic work $\dbar {\mathcal W} $ is independent of the state of the super-bath.

Note that the work $\dbar \tilde {\mathcal W}$ as given by Eq.~(\ref{dW-def-1}) depends on $\xv, \yv, \lambda, d \lambda$.  In stochastic thermodynamics, we keep track of dynamic evolution of $\xv$ but not of $\yv$.  Hence to obtain the differential work at the trajectory level of system variables, we need to average Eq.~(\ref{dW-def-1}) over the conditional equilibrium as given by Eq.~(\ref{p_Y-EQ-1}):
\ba
\dbar {\mathcal W} &=&
 \int_{\yv}  p^{\rm EQ}_{\Yv|\Xv}(\yv|\xv) \, 
\dbar \tilde{\mathcal W}
\nonumber\\
&=& \int_{\yv}  p^{\rm EQ}_{\Yv|\Xv}(\yv|\xv)  
\left(  d_{\lambda}{ H_{\Xv}} + d_{\lambda}{ H_{I}} \right).
\label{ddd-sss}
\ea
This equation and many analogous equations below are understood as volume integral of differential forms.  Be careful not to confuse the differential forms $\dbar \tilde{\mathcal W}, d_{\lambda}{ H_{\Xv}}$ etc with the volume measure $d^N \! y$ which is hidden in $\int_{\yv}$. 

Now, taking the $\lambda$ differential of Eq.~(\ref{Z_Y-def-0}), and further using Eq.~(\ref{H_I-constraint-1}), we find:
\be
 \int_{\yv}  p^{\rm EQ}_{\Yv|\Xv}(\yv|\xv)    d_{\lambda}{ H_{I}} = 0. 
 \label{useful-result-0}
 \ee
Hence Eq.~(\ref{ddd-sss}) reduces to 
\be
\dbar {\mathcal W} 
= d_{\lambda} H_{\Xv}
 = \frac{\partial H_{\Xv}}{\partial \lambda}d \lambda. 
\label{dW-def-2}
\ee
Hence, even though the interaction Hamiltonian $H_I$ may be tuned externally, the work $\dbar {\mathcal W}$ at trajectory level is nonetheless independent of $H_I$.  
 }
 
%Let us re-emphasize that $\dbar \tilde {\mathcal W}$ is the work at the trajectory level of the universe whereas $\dbar {\mathcal W}$ is the work at the level of system.  
%In thermodynamic theory of the system, we only keep track of the trajectory of the system  but not that of the universe.  Hence it is $\dbar {\mathcal W}$ which can measured and calculated.   } This definition of work Eq.~(\ref{dW-def-2}) is formally identical to that in weak coupling theory of stochastic thermodynamics.  

Taking the differential of Eq.~(\ref{H_XYZ-def}) and using Eq.~(\ref{dE-tot-lambda-0}), we obtain 
\be
 d H_{\Xv} = \dbar \tilde {\mathcal W} - d \left( H_{B} + H_{\Zv} \right). 
\label{dQ-def-path-0}
\ee
As in above, we take the average Eq.~(\ref{dQ-def-path-0}) over fluctuations of $\Yv\Zv$, which results in 
\ba
dH_{\Xv} &=&  \dbar {\mathcal W} + \dbar {\mathcal Q},  
\label{1st-law-path}\\
\dbar {\mathcal Q} &\equiv& - d \langle H_{B} 
+ H_{\Zv} \rangle_{\Yv\Zv},
\label{dQ-HB}
\ea
where $\langle \,\cdot \, \rangle_{\Yv\Zv}$ means average over $\Yv\Zv$, and $\dbar {\mathcal Q} $ is the differential heat at trajectory level of the system variables. 
Since $H_{\Xv}$ is defined as the fluctuating internal energy, and $ \dbar {\mathcal W} $ is the work at the trajectory level, Eq.~(\ref{1st-law-path}) can be interpreted as the first law at the trajectory level  if $\dbar {\mathcal Q} = - d \left( H_{B} + H_{\Zv} \right)$ is interpreted as the {\em heat at the trajectory level}.  
  Equation (\ref{dQ-HB}) then says that the heat $\dbar {\mathcal Q} $ is negative the average energy variation of the environment $\Yv\Zv$.  Such an interpretation of heat is exactly the same as that in the weak coupling stochastic thermodynamics.  

But the differential of fluctuating internal energy $dH_{\Xv} $ can be written as the sum of $d_{\lambda} H$ and $d_{\xv}H_{\Xv}$:
\be
dH_{\Xv} = d_{\lambda} H_{\Xv} + d_{\xv}H_{\Xv}. 
\label{dH_x-decomp}
\ee
Comparing this with Eq.~(\ref{1st-law-path}) we obtain an alternative expression for $\dbar {\mathcal Q}$: 
\be
\dbar {\mathcal Q} \equiv dH_{\Xv} - \dbar {\mathcal W}
= d_{\xv}H_{\Xv},
\label{dQ-def-path}
\ee
which must be equivalent to Eq.~(\ref{dQ-HB}).    It is attempting to rewrite $ d_{\xv}H_{\Xv}$ in terms of partial derivatives 
\be
d_{\xv}H_{\Xv} = \frac{\partial H_{\Xv}}{\partial \xv} d\xv. 
\label{d_xH_X}
\ee
\xing{This is however valid only if $\xv(t)$ is differentiable so that $d\xv$ is linear in $dt$.  In the limit of time-scale separation, we expect that a typical path of slow variables $\xv(t)$ becomes that of Brownian motion, which is everywhere continuous but non-differentiable.  As a consequence, $d\xv(t)$ scales as $\sqrt{dt}$ (Ito's formula) and we need to expand $d_{\xv}H_{\Xv} $ up to the second order in $d\xv$, if the product in RHS of Eq.~(\ref{d_xH_X}) is defined in Ito's sense.  We can also interpret the product  in RHS of Eq.~(\ref{d_xH_X}) in Stratonovich's sense.  Then Eq.~(\ref{d_xH_X}) remains valid even if $\xv(t)$ is a typical path of Brownian motion.  In this work, we shall not write $d_{\xv}H_{\Xv} $ in terms of partial derivatives, so that we do not need to worry about the issue of stochastic calculus. 

Note that the definitions of work and heat at trajectory level, Eqs.~(\ref{dW-def-2}) and (\ref{dQ-def-path}), are the same as those in the weak coupling theory.  
}

\subsection{Work and heat at ensemble level} 
\xing{
To obtain work and heat  at ensemble level, we need to average corresponding objects at trajectory level over (generally out-of-equilibrium) statistical distribution of dynamic trajectories of $\Xv$.  This is a rather nontrivial task.  Luckily, $ \dbar {\mathcal W}$ as given by Eq.~(\ref{dW-def-2}) is independent of $d\xv$.   Hence we do not need to know the pdf of $d\xv$, but only need to average Eq.~(\ref{dW-def-2}) over statistical distribution $p_{\Xv}(\xv, t)$ at time $t$, and obtain the differential work $\dbar W$ at ensemble level:
\be
\dbar W  = \int_{\xv} p_{\Xv} d_{\lambda} H_{\Xv}. 
\label{dW-ensemble-def}
\ee

Now we want to take ensemble average of heat Eq.~(\ref{dQ-def-path}), which does depend on $d\xv \equiv \xv(t + dt) - \xv(t)$, whose distribution is not encoded in the instantaneous distribution $p_{\Xv}(\xv, t)$.  A dynamic theory for $d\xv$, such as non-linear Langevin dynamics, would supply the necessary information.  This route was pursued in the complementary work~\cite{covariant-Thermodynamics-2021}.  Here we take a detour by studying the average of $d H_{\Xv}$. Let $p_{\Xv}(\xv, t)$ and $p_{\Xv}(\xv, t + dt)$ be the pdfs of $\xv$ at $t$ and at $t + dt$ respectively, and 
$d p_{\Xv}(\xv, t)$ the differential of $p_{\Xv}(\xv, t)$ as given by
 \ba
d p_{\Xv}(\xv, t) &=& p_{\Xv}(\xv, t + dt) -  p_{\Xv}(\xv, t) ,
\nonumber \\
&=& \frac{\partial p_{\Xv}(\xv, t)}{\partial t} dt. 
\ea
Let us calculate the differential of internal energy as given by Eq.~(\ref{E-ensemble-def}):
\be
d E_{\Xv}[p_{\Xv}] = d  \langle H_{\Xv} \rangle = d \int_{\xv} H_{\Xv}  p_{\Xv}.  
\ee
Since $\xv$ is integrated over in RHS, the differential $d$ is due to changes of $\lambda$ and of $p(\xv, t)$:
\be
d \langle  H_{\Xv} \rangle =  
\int_{\xv} (  d_{\lambda} H_{\Xv})  \, p_{\Xv}
+  \int_{\xv} H_{\Xv} \, d p_{\Xv}  .  
\label{1st-law-ensemble-0-1}
\ee
But the first term in RHS is just the work at ensemble level, as we defined in Eq.~(\ref{dW-ensemble-def}).  Hence the second term must be the heat at ensemble level:
\be
\dbar Q = \int_{\xv} H_{\Xv} d p_{\Xv}
= \langle \dbar {\mathcal Q} \rangle,  
\label{dQ-def-ensemble}
\ee
and Eq.~(\ref{1st-law-ensemble-0-1}) becomes the first law at the ensemble level:
\be
dE_{\Xv} [p_{\Xv}] = \dbar W + \dbar Q.  
\label{1st-law-ensemble-0}
\ee
The definitions of work and ensemble at ensemble level, Eqs.~(\ref{dW-ensemble-def}) and (\ref{dQ-def-ensemble}), are again the same as those in the weak coupling theory of stochastic thermodynamics.  }

%where $E_{\Xv},\dbar W $ are ensemble averages of $H_{\Xv},  \dbar {\mathcal W}$:
%\ba
%E_{\Xv} &=&  \int_{\xv} p_{\Xv} H_{\Xv}, 
%\label{E-ensemble-def}
%\ea 
%Now we can take differential of Eq.~(\ref{E-ensemble-def}):
%\ba
%d E_{\Xv} = d \int_{\xv} p_{\Xv} H_{\Xv}
% =  \dbar W + \int_{\xv} H_{\Xv} d p_{\Xv},
%\label{1st-law-ensemble}
%\ea
%where $\xv$ must be  treated as dummy variables, and $d p_{\Xv}$ is understood as due to the coarse-grained dynamics of the system variables.  Comparing this with Eq.~(\ref{1st-law-ensemble-0}) we obtain the definition of heat at ensemble level:
%\be
%\dbar Q = \int_{\xv} H_{\Xv} d p_{\Xv}
%= \langle \dbar {\mathcal Q} \rangle,  
%\label{dQ-def-ensemble}
%\ee
%where $\langle \, \cdot\, \rangle$ means averaging over slow variables.  In Ref.~\cite{covariant-Thermodynamics-2021} it was explicitly proved that Eq.~(\ref{dQ-def-ensemble}) is the ensemble average of Eq.~(\ref{dQ-def-path}) using Ito-Langevin theory.   

\section{Physical Meanings of Heat} 
\label{sec:heat}
In this section, we shall establish the connections between heat (both at trajectory level and at ensemble level) and entropy change of the environment, conditioned on the system state and possibly other thermodynamic variables.   We shall also discuss the physical meanings of Clausius inequality and total entropy production.  The results are again the same as those in the weak coupling theory, with the conditioning of slow variables properly taken into account.  

\subsection{Heat at trajectory level}
\xing{
The universe $\Xv\Yv\Zv$ is thermally closed, and evolves according to Hamiltonian dynamics with Hamiltonian given by Eq.~(\ref{H_XYZ-def}).   
Due to TSS, with $\xv$ fixed, the {\em environment} $\Yv\Zv$ is described by a micro-canonical ensemble with fixed energy.   We define Boltzmann entropy of environment as a function of its energy $E_{\Yv\Zv}$:
\ba
 & & S_{\Yv\Zv} (E_{\Yv\Zv}) \equiv \log \Omega_{\Yv\Zv} (E_{\Yv\Zv}) 
\nonumber\\
&\equiv& \log \int_{\yv,\zv}\, 
\delta(  H_{\rm Bath} + H_{\Zv} - E_{\Yv\Zv}), 
\quad \label{S_Y-def-1}
\ea
where $H_{\rm Bath}$ is defined in Eq.~(\ref{H_Bath-def}), and $\Omega_{\Yv\Zv} (E_{\Yv\Zv}) $ is the area of $\Yv\Zv$ hyper-surface with constant bath energy $E_{\Yv\Zv}$.    Note that $S_{\Yv\Zv} (E_{\Yv\Zv})$ generally also depends on $\xv, \lambda, \beta$ parametrically through $H_{\rm Bath}$.   We shall however not explicit display the parameters $\xv, \lambda, \beta$, in order not to make the notations too cluttered.  Strictly speaking, $S_{\Yv\Zv} (E_{\Yv\Zv})$ is {\em the Boltzmann entropy of the environment conditioned on $\Xv = \xv$}.   
 }
  
% $ \Omega_{\Yv} (E_{\Yv})  = e^{S_{\Yv} (E_{\Yv})}$ is related to $Z_{\Yv}(\beta)$ defined in Eq.~(\ref{Z_Y-def}) via a Laplace transform.  Hence, given $E_{\Yv}$, $ \Omega_{\Yv} (E_{\Yv}) , S_{\Yv} (E_{\Yv})$ are also independent of $\xv$ and $\lambda$. 

 Suppose in the initial state the system is at $\xv$ with external parameter $\lambda$, and the universe $\Xv\Yv\Zv$ has total energy $E_{\Xv\Yv\Zv} $.  The energy of the environment is then $E_{\Yv\Zv} = E_{\Xv\Yv\Zv} - H_{\Xv}$.     In the final state the system is at $\xv + d \xv$ with external parameter $\lambda + d \lambda$, and the universe  has total energy $E_{\Xv\Yv\Zv} + \dbar {\mathcal W}$, with $ \dbar {\mathcal W}$ given by Eq.~(\ref{dW-def-2}).  (Recall that the work is defined as the change of total energy.)  The energy of the environment in the final state is then $E_{\Yv\Zv}' = E_{\Xv\Yv\Zv} + \dbar {\mathcal W}   - H_{\Xv} - d H_{\Xv}$, where $d H_{\Xv}$ is given by Eq.~(\ref{dH_x-decomp}).   The Boltzmann entropies of the environment in the initial and final states are hence respectively:
\begin{subequations}
\ba
S_{\Yv\Zv}(E_{\Yv\Zv}) &=& S_{\Yv\Zv}(E_{\Xv\Yv\Zv} - H_{\Xv} ),
 \label{S'-1-0}\\
S_{\Yv\Zv} (E_{\Yv\Zv}')  &=& 
S_{\Yv\Zv}\Big( E_{\Xv\Yv\Zv} + \dbar {\mathcal W} 
 - H_{\Xv} - d H_{\Xv} \Big). 
 \nonumber\\
 \label{S'-1}
\ea
\end{subequations}
Note that $E_{\Xv\Yv\Zv}$ is much larger than $d H_{\Xv}, \dbar {\mathcal W} $, because the size of super-bath is much larger than $\Xv\Yv$.   Expanding Eq.~(\ref{S'-1}) in terms of $\dbar {\mathcal W}$ and $d H_{\Xv}$ and subtracting from it Eq.~(\ref{S'-1-0}), we obtain:
\ba
dS_{\Yv\Zv}(E_{\Yv\Zv}) &=&S_{\Yv\Zv} (E_{\Yv\Zv}') 
- S_{\Yv\Zv}(E_{\Yv\Zv})
\nonumber\\
 &=& \beta \left( \dbar {\mathcal W}  - d H_{\Xv} \right)
= - \beta  \dbar {\mathcal Q} ,   \quad 
 \label{dS_Y-1}
% -  d_{\lambda} H_{\Xv} \right) - \beta d_{\xv} H_{\Xv}  
%\nonumber\\
%&=& - \beta d_{\xv} H_{\Xv} (\xv; \lambda),
\ea 
where $\beta = \partial S_{\Yv\Zv}/\partial E_{\Yv\Zv}$ is the inverse temperature.
Further using  Eq.~(\ref{dQ-def-path}), we find
\ba
 - \beta  \dbar {\mathcal Q} = dS_{\Yv\Zv}(E_{\Yv\Zv}),
\ea 
which establishes the connection between the differential heat $\dbar {\mathcal Q}$ at the level of system trajectory and the differential of environment Boltzmann entropy $dS_{\Yv\Zv}(E_{\Yv\Zv})$ conditioned on $\Xv = \xv$.  
\xing{
\subsection{Heat at ensemble level, and total entropy production}
\label{sec:Q-ensemble}
Recall that $\Xv\Yv$ is in contact with a much larger super-bath $\Zv$, and that $\Yv$ is always in conditional equilibrium.  If the system is in a non-equilibrium state $p_{\Xv}(\xv)$, the joint pdf of $\Xv\Yv$ is given by
\be
p_{\Xv\Yv}(\xv, \yv) = p_{\Xv}(\xv)\, p^{\rm EQ}_{\Yv|\Xv}(\yv |\xv),
\label{p_XY-cond-EQ}
\ee
where $p^{\rm EQ}_{\Yv|\Xv}(\yv |\xv)$ is given in Eq.~(\ref{p_Y-EQ-1}).   The non-equilibrium free energy for the system is already defined in Eq.~(\ref{F_X-def}).  Let us similarly define the non-equilibrium free energy of the combined system $\Xv\Yv$: 
\ba
F_{\Xv\Yv}[p_{\Xv\Yv}] &\equiv& \int_{\xv,\yv} p_{\Xv\Yv} \left( 
H_{\Xv\Yv} + T \log p_{\Xv\Yv}\right). 
\label{F_XY-def}
\ea
For $\Xv\Yv$, there is no difference between Hamiltonian and Hamiltonian of mean force, since $\Xv\Yv$ is in weak interaction with $\Zv$.   Substituting Eq.~(\ref{p_XY-cond-EQ}) into Eq.~(\ref{F_XY-def}), and using Eqs.~(\ref{H_Bath-def}) and (\ref{F-Z_Y}), we obtain
\be
F_{\Xv\Yv}[p_{\Xv\Yv}]  = F_{\Xv}[p_{\Xv}] + F_{\Yv}(\beta).
\label{F_XY-F_X-Z_Y}
\ee
which says that $F_{\Xv\Yv}[p_{\Xv\Yv}]$ and $F_{\Xv}[p_{\Xv}]$ differ only by an additive constant $F_{\Yv}(\beta)$, which is, according to Eq.~(\ref{H_I-constraint-1}),  independent of $\lambda, \xv$, and hence need to be worried about when we study non-equilibrium processes.  Equation (\ref{F_XY-F_X-Z_Y}) is a non-equilibrium generalization of Eq.~(\ref{F_XY-canonical}).   

Let us now consider variations of $\lambda$ and $p_{\Xv}$, and study the resulting variation of free energies.  Taking the differential of Eq.~(\ref{F_X-def}), we obtain:
\be
dF_{\Xv}[p_{\Xv}]  = \dbar W + \dbar Q - T d S_{\Xv}[p_{\Xv}],
\ee 
where $\dbar W, \dbar Q$ are work and heat at ensemble level, given respectively in Eqs.~(\ref{dW-ensemble-def}) and (\ref{dQ-def-ensemble}).   We can rewrite this result into
\ba
 d S_{\Xv}[p_{\Xv}] - \beta \dbar Q 
= \beta \left( \dbar W -   dF_{\Xv}[p_{\Xv}] \right).  
\label{SQ-WF}
\ea
We can also do the similar thing on $dF_{\Xv\Yv}[p_{\Xv}] $, and obtain an analogous result:
\ba
 d S_{\Xv\Yv}[p_{\Xv\Yv}] - \beta \dbar Q_{\Xv\Yv} 
= \beta \left( \dbar W_{\Xv\Yv} -   dF_{\Xv\Yv}[p_{\Xv\Yv}] \right),
\nonumber\\
\label{SQ-WF-XY}
\ea
where $ \dbar W_{\Xv\Yv} , \dbar Q_{\Xv\Yv}  $ are the work and heat at ensemble level of $\Xv\Yv$:
\ba
 \dbar W_{\Xv\Yv} &=& \int_{\xv,\yv} p_{\Xv\Yv} d_{\lambda} H_{\Xv\Yv}, 
 \label{dW_XY}\\
  \dbar Q_{\Xv\Yv} &=& \int_{\xv,\yv} d p_{\Xv\Yv}\, H_{\Xv\Yv}. 
   \label{dQ_XY}
\ea

%Note that $\Xv\Yv$ is weakly coupled to $\Zv$, and hence $H_{\Xv\Yv}$ is also the HMF of $\Xv\Yv$.  

Using Eqs.~(\ref{p_XY-cond-EQ}) and (\ref{useful-result-0}) in Eq.~(\ref{dW_XY}), we see that 
\ba
\dbar W_{\Xv\Yv} = \dbar W.
\ea 

Taking the differential of Eq.~(\ref{F_XY-F_X-Z_Y}) we find
\be
dF_{\Xv}[p_{\Xv}]  =  dF_{\Xv\Yv}[p_{\Xv\Yv}] . 
\ee
Combining the preceding two equations with Eqs.~ (\ref{SQ-WF}) and (\ref{SQ-WF-XY}), we find 
\ba
&&  d S_{\Xv\Yv}[p_{\Xv\Yv}] - \beta \dbar Q_{\Xv\Yv} 
=  \beta \left( \dbar  W_{\Xv\Yv} -   dF_{\Xv\Yv}[p_{\Xv\Yv}] \right).
 \nonumber\\
&=&  d S_{\Xv}[p_{\Xv}] - \beta \dbar Q 
= \beta \left( \dbar W -   dF_{\Xv}[p_{\Xv}] \right).  
 \label{xx--xxx-0} 
\ea

Now recall $\Xv\Yv$ is weakly coupled to the super-bath $\Zv$, and hence the weak coupling theory of stochastic thermodynamics is applicable.  It tells us that $ d S_{\Xv\Yv}[p_{\Xv\Yv}] - \beta \dbar Q_{\Xv\Yv} $ is positive definite and can be interpreted as the change of total entropy of the universe $\Xv\Yv\Zv$.  Equation (\ref{xx--xxx-0}) then says that the total entropy production is the same, whether we calculate it using the dynamic theory of $\Xv\Yv$ or using the reduced theory $\Xv$ alone.  If we understand the dynamic theory of $\Xv$ as a consequence of coarse-graining of the $\Xv\Yv$ dynamics, then Eq.~(\ref{xx--xxx-0}) says that entropy production is invariant under coarse-graining, as long as the fast variables remain in conditional equilibrium.  A similar result was obtained by Esposito~\cite{Esposito-2012} in the setting of master equation dynamics.  

Furthermore, assuming that $\Xv\Yv$ evolves according to Langevin dynamics (which follows if the dynamics of $\Zv$ much faster than that of $\Xv\Yv$), the Clausius inequality can be proved using the Langevin dynamics $ d S_{\Xv\Yv}[p_{\Xv\Yv}] - \beta \dbar Q_{\Xv\Yv} \geq 0$.  Hence we have
\ba
  dS_{\Xv\Yv\Zv} &=&  d S_{\Xv\Yv}[p_{\Xv\Yv}] 
  - \beta \dbar Q_{\Xv\Yv}
  \nonumber\\
  &=&  \beta \left( \dbar  W_{\Xv\Yv} 
  -   dF_{\Xv\Yv}[p_{\Xv\Yv}] \right) \geq 0. 
 \label{xx--xxx} 
\ea

Combining Eqs.~(\ref{xx--xxx}) with (\ref{xx--xxx-0}), we finally obtain
\ba
  dS_{\Xv\Yv\Zv} &=&   d S_{\Xv}[p_{\Xv}] - \beta \dbar Q
 \nonumber\\
 &=& \beta \left( \dbar W -   dF_{\Xv}[p_{\Xv}] \right)
 \geq 0, 
 \label{dS_tot-1}
\ea
which not only establishes the Clausius inequality, but also says that the physical meaning of $d S_{\Xv}[p_{\Xv}] - \beta \dbar Q$ is indeed the variation of total entropy of the universe. 

It is interesting to rewrite Eq.~(\ref{dS_tot-1}) into
 \ba
 - \beta \dbar Q &=& d\left(S_{\Xv\Yv\Zv}[p_{\Xv\Yv\Zv}] 
 - S_{\Xv}[p_{\Xv}]  \right)
\nonumber\\
&=& d S_{\Yv\Zv|\Xv}. 
\label{dS-tot-dS-dQ-2}
 \ea
Hence $ - \beta \dbar Q$ is the differential of $S_{\Yv\Zv|\Xv}$, the conditional Gibbs-Shannon entropy of $\Yv\Zv$ given the system state $\Xv$.  
 }

%For the weak coupling case, the correlation between $\Xv$ and $\Yv$ is negligible, and the conditioning can be removed, which leads to the familiar result $\dbar Q = - T d S_{\Yv}$.  

%With the assumption of TSS, we expect that the dynamics of slow variables obeys nonlinear Langevin theory.  It then can be proved that, with work and heat defined by Eqs.~(\ref{dW-ensemble-def}) and (\ref{dQ-def-ensemble}), the RHS of Eq.~(\ref{SQ-WF}) is positive definite~\cite{}.    Combination of this inequality and Eq.~(\ref{SQ-WF}) then implies the Clausius inequality.  In this sense, the validity of the second law can be traced back to the ubiquity of TSSs, and to our incapability, as macroscopic creatures, of resolving the fast-scale details.  

\vspace{-3mm}

\xing{
\section{Comparison with Other Theories}
\label{sec:comparison}
In this section, we provide a detailed comparison between the present work and several previous influential works on  strong coupling thermodynamics.  First of all, we list all major formulas of our theory in the central column of Table I.  These formulas are identical to those of the weak coupling stochastic thermodynamic theory, with $H_{\Xv}$ understood as the Hamiltonian of mean force.   In the weak coupling limit, $H_{\Xv}$ simply becomes the bare Hamiltonian of the system. 

\begin{table*}[htp!]
\setlength{\extrarowheight}{7pt}
\begin{tabular}{ l  l l }
 % \hline \hline \
    \bf \shortstack{TM Quantities\\and Laws}  \quad\quad \quad
    & \bf  \shortstack{Weak Coupling theory \\ and the Present Theory} & \bf Seifert's Strong Coupling theory
\vspace{1mm}\\  \hline\hline 
Fluctuating Internal Energy \quad\quad  & $H_{\Xv}$
& $ \tilde H_{\Xv} \equiv \partial_{\beta} \beta H_{\Xv}$  
\vspace{1mm} \\ \hline  
 Internal Energy & $E_{\Xv}[p_{\Xv}] = \int_{\xv} p_{\Xv} H_{\Xv}$
& $\tilde E_{\Xv}[p_{\Xv}] = \int_{\xv} p_{\Xv}  \partial_{\beta} \beta H_{\Xv}$ 
\vspace{1mm} \\ \hline  
 Entropy & $S_{\Xv}[p_{\Xv}] = - \int_{\xv} p_{\Xv} \log p_{\Xv} $  \quad\quad 
 &  $\tilde S_{\Xv}[p_{\Xv}] = \int_{\xv} p_{\Xv} ( - \log p_{\Xv} + \beta^2 \partial_{\beta} H_{\Xv}) $ 
\vspace{1mm} \\ \hline  
Free energy & $F_{\Xv} = E_{\Xv} - TS_{\Xv}$ & $F_{\Xv} = \tilde E_{\Xv} - T \tilde S_{\Xv}= E_{\Xv} - TS_{\Xv}$
\vspace{1mm} \\ \hline  
% TM relations & \xing{$E \neq \frac{\partial \beta F}{\partial \beta}, 
% \quad S \neq - \beta^2 \frac{\partial F}{\partial \beta} $} 
% & $ \tilde E = \frac{\partial \beta F}{\partial \beta}, \quad
% \tilde S = - \beta^2 \frac{\partial F}{\partial \beta} $ 
%\vspace{1mm} \\ \hline  
 Work  at trajectory level & $\dbar {\mathcal W} =  d_{\lambda} H_{\Xv}$ 
 &$\dbar {\mathcal W} =  d_{\lambda} H_{\Xv}$ 
\vspace{1mm} \\ \hline  
Heat at trajectory level & $\dbar {\mathcal Q} = d_{\xv}  H_{\Xv}  $ 
 &$\dbar \tilde {\mathcal Q} = d_{\xv}  H_{\Xv}  
 + \beta \partial_{\beta} d_{\xv} H_{\Xv}
 + \beta \partial_{\beta} d_{\lambda} H_{\Xv}$ 
\vspace{1mm} \\ \hline  
1st Law at trajectory level & $dH_{\Xv} = \dbar {\mathcal W} + \dbar  {\mathcal Q}$ & $d \tilde H_{\Xv} = \dbar {\mathcal W} + \dbar  \tilde {\mathcal Q}$
\vspace{1mm} \\ \hline  
    Work  at ensemble level & $\dbar W = \int_{\xv} p_{\Xv}\, d_{\lambda} H_{\Xv}$ 
 & $\dbar W = \int_{\xv} p_{\Xv} \, d_{\lambda} H_{\Xv}$ 
\vspace{1mm} \\ \hline  
     Heat   at ensemble level & $\dbar { Q} =  \int_{\xv} H_{\Xv} \, dp_{\Xv}$ 
 & $\dbar \tilde { Q} = \int_{\xv} \left( 
 (\partial_{\beta} \beta H_{\Xv}) dp_{\Xv}
  + \beta \partial_{\beta} (d_{\lambda} H_{\Xv} )  p_{\Xv} \right)$ 
\vspace{1mm} \\ \hline  
1st Law at ensemble level & $d E_{\Xv} = \dbar { W} + \dbar  { Q}$ & $d \tilde E_{\Xv} = \dbar { W} + \dbar  \tilde { Q}$
\vspace{1mm} \\ \hline  
2nd law (Clausius ineq.) & $dS^{\rm tot} = dS_{\Xv} - \beta \dbar Q = \beta (\dbar W - dF_{\Xv})\geq 0$ 
 \quad\quad \quad & 
 $dS^{\rm tot} = d\tilde S_{\Xv} - \beta \dbar \tilde Q =  \beta (\dbar W - dF_{\Xv}) \geq 0$ 
\vspace{1mm} \\ \hline  
Crooks FT & $p_{F}(W) = p_R(-W) e^{\beta (W - \Delta F_{\Xv})}$
 & $p_{F}(W) = p_R(-W) e^{\beta (W - \Delta F_{\Xv})}$
\vspace{1mm} \\ \hline  
Jarzynski inequality& $\langle e^{-\beta W} \rangle = e^{-\beta \Delta F_{\Xv}}$
 & $\langle e^{-\beta W} \rangle = e^{-\beta \Delta F_{\Xv}}$
  \vspace{1mm}  \\ \hline  \hline
      \label{table-I}
      \end{tabular}
\vspace{-5mm}
\caption{Major formulas of thermodynamics and stochastic thermodynamics.  According to the present theory, formulas in the central column are applicable both in the weak coupling regime and in the strong coupling regime, with $H_{\Xv}$ the Hamiltonian of mean force.  The formulas in Seifert's theory are shown in the right column, which are substantially more complex.  The differences disappear in the weak coupling limit, where $H_{\Xv}$ reduces to the bare system Hamiltonian which is independent of $\beta$.  }
 %\vspace{-3mm} 
\end{table*}

In the theory developed by Seifert~\cite{Seifert-2016-strong-coupling} and critically evaluated by Talkner and H\"anggi~\cite{Hanggi-2016-strong-coupling}, the equilibrium free energy of a strongly coupled system is defined in terms of HMF $H_{\Xv}$ as
\ba
F_{\Xv} &=& - T \log Z_{\Xv} = 
-T \log \int_{\xv} e^{-\beta H_{\Xv}(\xv; \lambda, \beta)},  
\label{S-E-tilde-EQ-F}
\ea
which is the same as Eq.~(\ref{F-Z_X}).   The equilibrium internal energy and entropy are { defined}  as:
\begin{subequations}
\ba
\tilde E_{\Xv}  &\equiv& 
 \frac{{\partial \beta F_{\Xv}   }}{\partial \beta}
 = \int  p_{\Xv}^{\rm EQ} (  H_{\Xv}  
 + \beta \, \partial_{\beta} H_{\Xv}) , 
 \label{E-S-F-derivatives-new-E} \\ 
%=  - \frac{1}{Z }  \frac{{\partial  }}
%{\partial \beta} \int  e^{-\beta H }, \\
\tilde S_{\Xv}  &\equiv&  
- \beta ^2 \frac{\partial F_{\Xv}  } {\partial \beta}
=  \int  p_{\Xv}^{\rm EQ} 
\left( - \log  p_{\Xv}^{\rm EQ}
+ \beta^2 \partial_\beta H_{\Xv} 
\right). 
\nonumber\\
\label{E-S-F-derivatives-new-S}
\ea
\label{E-S-F-derivatives-new}
\end{subequations}
such that $F_{\Xv} = \tilde E_{\Xv} - T \tilde S_{\Xv}$ remains valid.  (We use $\tilde A$ to denote thermodynamic quantity in Seifert's theory if it is different from the corresponding quantity $A$ in our theory.)  Note that in our theory, energy and entropy are defined by Eqs.~(\ref{reduced-X-thermodynamics}).      

%Using Eq.~(\ref{S-E-tilde-EQ-F}),  Eqs.~(\ref{E-S-F-derivatives-new}) can be  rewritten as
%\begin{subequations}
%\ba
%\tilde E_{\Xv}  &=& \int  p_{\Xv}^{\rm EQ}(\xv) (  H_{\Xv} 
% + \beta \, \partial_{\beta} H_{\Xv}) , \\
%\tilde S_{\Xv}  &=& \int  p_{\Xv}^{\rm EQ} (\xv)
%\left( - \log  p_{\Xv}^{\rm EQ}(\xv) 
%+ \beta^2 \partial_\beta H_{\Xv} 
%\right). \quad \quad
%\ea
%\label{S-E-tilde-EQ}
%\end{subequations}

In review of the results obtained in Sec.~\ref{sec:joint-EQ}, the following thermodynamic relations hold for $\Xv\Yv$:
\ba
 E_{\Xv\Yv}  &=& 
 \frac{{\partial \beta F_{\Xv\Yv}   } } {\partial \beta}, \quad
 S_{\Xv\Yv}   = - \beta ^2 \frac{\partial F_{\Xv\Yv}  } {\partial \beta}.
\label{E-S-F-derivatives-new-XY}
\ea
The free energy, energy, and entropy of the bath are then defined as
\begin{subequations}
\ba
\tilde F_{\Yv} &=& F_{\Xv\Yv} -  F_{\Xv} = F_{\Yv} = F_{\Yv}^0, \\
\tilde E_{\Yv} &=& E_{\Xv\Yv} - \tilde E_{\Xv}, \\
\tilde S_{\Yv} &=& S_{\Xv\Yv} - \tilde S_{\Xv},
\ea
\end{subequations}
where used was Eq.~(\ref{Z_Y-def}).  Combining theses with Eqs.~(\ref{E-S-F-derivatives-new-XY}) and (\ref{E-S-F-derivatives-new}), we see that the bath energy and entropy in Seifert's theory satisfy 
\ba
\tilde E_{\Yv}  &=& 
 \frac{{\partial \beta F_{\Yv}^0   } } {\partial \beta}, \quad
\tilde S_{\Yv}   = - \beta ^2 \frac{\partial F_{\Yv}^0 } {\partial \beta},
\label{E-S-F-derivatives-new-bath}
\ea
where $F^0_{\Yv}$ is the free energy of the bare bath, with the interaction switched off.  These results show that in Seifert's theory, the interaction energy and correlation are completely relegated to the system.   By contrast, in our theory, interaction energy and correlation are completely relegated to the bath, if we interpret $H_{\Xv}$ as the system Hamiltonian. 

Seifert further bootstrap Eqs.~(\ref{E-S-F-derivatives-new}) to the non-equilibrium case, and define fluctuating internal energy $\tilde H $, non-equilibrium internal energy $\tilde E [p_{\Xv} ]$, and non-equilibrium entropy $ \tilde S [p_{\Xv}  ] $ as follow:
\begin{subequations}
\ba
\tilde H_{\Xv}  &\equiv& H_{\Xv} 
 + \beta \, \partial_{\beta} H_{\Xv} 
 = \partial_{\beta } (\beta H_{\Xv}) , \\
\tilde E_{\Xv} [p_{\Xv} ]   &=& \int_{\xv}  p_{\Xv} (  H_{\Xv} 
 + \beta \, \partial_{\beta} H_{\Xv} ) , \\
 \tilde S_{\Xv} [p_{\Xv}  ]  &\equiv&  \int_{\xv}  p_{\Xv} 
 \left( - \log p_{\Xv}  
 +  \beta^2 \partial_\beta H_{\Xv} \right), 
\ea
\label{S-E-tilde-NEQ}
\end{subequations} 
The differential of entropy is then given by
\ba
d \tilde S_{\Xv} &=& - \int_{\xv} \log p_{\Xv} d p_{\Xv}  
   + \int_{\Xv} p_{\Xv} \beta^2 \partial_{\beta} d_{\lambda} H_{\Xv}.  
   \label{dS-tilde-def}
\ea
The non-equilibrium free energy is defined as
\ba
\tilde  F_{\Xv} [p_{\Xv} ]  &\equiv&
 \tilde E_{\Xv} [p_{\Xv}  ]     - \tilde S_{\Xv}  [p_{\Xv}  ]  
\nonumber\\
&=& \int_{\xv} p_{\Xv} \left( 
H_\Xv + T \log p_\Xv \right) 
\nonumber\\
&=& F_{\Xv} [p_{\Xv} ] ,
\ea
which is the same as that of our theory, Eq.~(\ref{F_X-def}).  

The work at trajectory level and ensemble level are defined in terms of change of total energy:
\ba
\dbar {\mathcal W} &\equiv& d H_{\Xv\Yv \Zv} = d_{\lambda} H_{\Xv}^0
= d_{\lambda} H_{\Xv}, \\
\dbar { W} &\equiv& \int_{\xv} p_{\Xv} \, d_{\lambda} H_{\Xv},
\ea
which are identical to our definitions.  The heat at trajectory level is then defined to satisfy the first law:
\ba
\dbar \tilde {\mathcal Q} &\equiv& d \tilde H_{\Xv} - \dbar {\mathcal W}
 \nonumber\\
&=& d_{\xv}  H_{\Xv}  
 + \beta \partial_{\beta} d_{\xv} H_{\Xv}
 + \beta \partial_{\beta} d_{\lambda} H_{\Xv},
 \\
\dbar \tilde { Q} &\equiv& \int_{\xv} \left( 
 (\partial_{\beta} \beta H_{\Xv}) dp_{\Xv}
  + \beta \partial_{\beta} (d_{\lambda} H_{\Xv} )  p_{\Xv} \right).  
  \label{dQ-tilde-def}
\ea
The LHS of Clausius inequality can be calculated:
\ba
d \tilde S_{\Xv} - \beta \dbar \tilde Q &=& d  S_{\Xv} - \beta \dbar  Q
= \beta \left( \dbar W - dF_{\Xv} \right)
\nonumber\\
&=&- \int_{\xv} \left( \log p_{\Xv} + \beta H_{\Xv}
\right) dp_{\Xv},
\label{equivalence-dS-tot}
\ea
which is again the same as in our theory.  As a consequence, the first and second laws of thermodynamics in Seifert's theory are equivalent to those in our theory.  This means that these two theories are equivalent to each other, even though they use different definitions of internal energy, entropy, and heat.  Major formulas of Seifert's theory are displayed in the right column of Table I. 

%These relations, which are called by Hanggi and Talkner as ``thermodynamic consistency conditions'', can be deemed the definitions of system energy and entropy in Seifert's theory.  Seifert's definition of heat is also different from ours.  The variation of total entropy of the universe $dS - \beta \dbar Q$, however, are the same in two theories.  

Hanggi and Talkner~\cite{Hanggi-2016-strong-coupling,Hanggi-2020-Colloquium} accept the definitions of equilibrium thermodynamic quantities, Eqs.~(\ref{E-S-F-derivatives-new}) .  Yet they argue that the non-equilibrium thermodynamic quantities cannot be uniquely determined from their equilibrium versions, which is of course valid.  They also argue that the Hamiltonian of mean force cannot be uniquely determined from the equilibrium distribution of system variables alone~\footnote{See, however, the recent work by Strasberg and Esposito on this issue~\cite{Strasberg-2020-HMF}.}.  They further discuss more serious ambiguities associated with the definition of non-equilibrium work for quantum systems. 

Jarzynski~\cite{Jarzynski-2017-strong-coupling} develops a more comprehensive (and hence more complex) theory for strong coupling thermodynamics, and systematically discusswa the definitions of internal energy, entropy, volume, pressure, enthalpy, and Gibbs free energy.   Using a pebble immersed in a liquid as a metaphor, he establishes his formalism around the concept of volume, whose definition is somewhat arbitrary.   All other thermodynamic variables are uniquely fixed by thermodynamic consistency once the system volume is (arbitrarily) defined. Jarzynski further shows that Seifert's theory is a special case of his framework, i.e., the ``partial molar representation''.  He discusses in great detail the ``bare representation'', where the system enthalpy coincides with HMF.  The total entropy production is however the same in both representations.    The heat and work in the bare representation are formally identical to those in our theory.   We note that for many small systems, volume or pressure is seldom controlled.  It is then unnecessary to distinguish energy from enthalpy, or Helmholtz free energy from Gibbs free energy.   

% Note however Jarzynski does not assume TSS.  Hence it is not possible for him to assign physical meanings to differential heat, such as what we have done in Eqs.~(\ref{dS_Y-1}) and (\ref{dS-tot-dS-dQ-2}).   

In all works discussed above, the interaction Hamiltonian $H_I$ is assumed to be independent of the external parameter $\lambda$, whereas time-scale separation is not assumed.  As a consequence, it is possible to prove the integrated Clausius inequality $\Delta S - \beta Q \geq 0$ for a finite process, but not possible to prove the differential Clausius inequality $d S - \beta \dbar Q \geq 0$ for every infinitesimal evolution step in the process.   Barring the issues of TSS and of $\lambda$ dependence of the interaction Hamiltonian $H_I$, our theory can be understood as a simplification of Jarzynski's bare representation, with  HMF and free energy playing the role of enthalpy and Gibbs free energy.    

Strasberg and Esposito~\cite{Esposito-strong-coupling-2017} study the consequences of TSS in the settings both of master equation theory and of Hamiltonian dynamics. For master equation theory, using the conditional equilibrium nature of the fast variables, they show that a reduced theory of slow variables can be derived once the fast variables are averaged out.  Note, however, the heat and internal energy in their reduced theory pertain to the original system consisting of both slow and fast variables, see Eq.~(33)-(35) of Ref.~\cite{Esposito-strong-coupling-2017}.  As a  consequence, these quantities do not have a finite limit as the dimension of fast variables goes to infinite.   For  Hamiltonian dynamics, Strasberg and Esposito {\em propose} a definition of total entropy production as the relative entropy, and show that, with TSS, it is equivalent to that in Seifert's theory, which is also equivalent to entropy production in our theory, as we have demonstrated in Eq.~(\ref{equivalence-dS-tot}).  By this, they confirm the consistency of Seifert's strong coupling theory. 

By contrast, in the present work, we use TSS to carry out a different decomposition of Hamiltonian as discussed in Sec.~\ref{sec:H-decomp}.  This leads to a remarkable situation where all formulas of the weak coupling theory of stochastic thermodynamics remain applicable even in the strong coupling regime.  These formulas are significant simpler than those in Seifert's strong coupling theory.   For a comparison, see Table I.  

The differences between the present theory and Seifert's theory are however not completely notational.  Consider a ``fast'' slow process with time duration $dt$ where $\lambda$ changes by $d \lambda$.  It is slow enough so that the bath remains in conditional equilibrium, and our stochastic thermodynamic theory remains applicable. Yet it is also fast enough so that the distribution $p_{\Xv}$ barely changes. Such a process can always be realized if TSS is satisfied.     Hence we have $d_{\lambda} H_{\Xv} \neq 0$, but $dp_{\Xv} = 0$. According to the present theory, then both $dS_{\Xv}$ and $\dbar Q$ vanish, and hence the variation of total entropy $dS_{\Xv} - \beta \dbar Q$ also vanishes.  Now in Seifert's theory,  $d \tilde S_{\Xv}$ and $\dbar \tilde Q$ are given respectively by Eqs.~(\ref{dS-tilde-def}) and (\ref{dQ-tilde-def}).  Neither of these two vanishes even if $dp_{\Xv} = 0$,  yet the variation of the total entropy $d \tilde S_{\Xv} - \beta \dbar\tilde  Q$ does vanish.  This means that in Seifert's theory there is an exchange of entropy between the system and bath even though $p_{\Xv}$ remains unchanged.  While this does not violate the second law of thermodynamics, it does contradict the common intuition about  entropy as measure of multitude of system states: It is very strange if the pdf of system variable stay unchanged, yet the system entropy changes suddenly!  From this perspective, the present theory is more natural and intuitive. 
}

%\vspace{-3mm}
\section{Conclusion}   
\label{sec:conclusion}
%\vspace{-3mm}

In this work, we have demonstrated that the usual theory of strong coupling thermodynamics and stochastic thermodynamics, which is based on the assumption of weak coupling between the system and its environment, can be made applicable in the strong coupling regime, if we define the Hamiltonian of mean force as the system Hamiltonian.  Our result is consistent with previous theories by various authors, in the sense that the first and second laws in different theories are mathematically equivalent.   Overall, the present work can be understood as a re-interpretation,  synthesis, and simplification of various previous theories of strong coupling stochastic thermodynamics.  

In a future work, we will conduct a systematic study of coarse-graining process, i.e. integrating out fast variables to obtain an effective dynamic theory for slow variables, with the ratio of time scales between the slow and fast variables treated as a small parameter.  If this ratio is small but nonzero, there should be slight deviation of fast variables from conditional equilibrium.  We shall analyze how this deviation leads to modification of dissipation in the dynamics of slow variables.  We shall also extend our theory to quantum case, and develop a thermodynamic theory for small open quantum systems strongly coupled to environment.

  %We are reasonably optimistic that with the assumption of time-scale separation, there exists a comprehensive and coherent theory for non-equilibrium statistical mechanics, on a par with Gibbs' ensemble theory for equilibrium statistical physics.  
 
 %We expect the bulk of the present theory remains applicable for quantum systems as well. 

%
\vspace{2mm} 
%
%\section{acknowledgement}

X.X. acknowledges support from NSFC grant \#11674217  as well as Shanghai Municipal Science and Technology Major Project (Grant No.2019SHZDZX01).
Z.C.T. acknowledges support from NSFC grant \#11675017.    X.X. is also thankful to additional support from a Shanghai Talent Program.


\begin{thebibliography}{99}

\bibitem{seifert2005entropy}
Seifert, Udo.  ``Entropy production along a stochastic trajectory and an integral fluctuation theorem.'' Physical review letters, 2005, 95(4): 040602.

\bibitem{Seifert-review}
Seifert, Udo. ``Stochastic thermodynamics, fluctuation theorems and molecular machines.'' Reports on progress in physics 75.12 (2012): 126001.

\bibitem{Jarzynski-review}
Jarzynski, Christopher. ``Equalities and inequalities: Irreversibility and the second law of thermodynamics at the nanoscale.'' Annu. Rev. Condens. Matter Phys. 2.1 (2011): 329-351.

%%%%%% 

\bibitem{Jarzynski-2004-HMF}
Jarzynski, Chris. ``NonEQ work theorem for a system strongly coupled to a thermal environment.'' Journal of Statistical Mechanics: Theory and Experiment 2004.09 (2004): P09005.

\bibitem{Seifert-2016-strong-coupling}
Seifert, Udo. ``First and second law of thermodynamics at strong coupling.'' Physical review letters 116.2 (2016): 020601.

\bibitem{Hanggi-2016-strong-coupling}
Talkner, Peter, and Peter H\"anggi. ``Open system trajectories specify fluctuating work but not heat.'' Physical Review E 94.2 (2016): 022143.

\bibitem{Jarzynski-2017-strong-coupling}
Jarzynski, Christopher. ``Stochastic and macroscopic thermodynamics of strongly coupled systems.'' Physical Review X 7.1 (2017): 011008.

\bibitem{Hanggi-2020-Colloquium}
Talkner, Peter, and Peter Hänggi. 
``{\it Colloquium:} Statistical Mechanics and Thermodynamics at Strong Coupling:  Quantum and Classical.''  Rev. Mod. Phys. 92, 041002 (2020). 

%%%%%%  

\bibitem{Strasberg-2016-strong-coupling}
Strasberg, P., Schaller, G., Lambert, N., \& Brandes, T. (2016). ``Non equilibrium  thermodynamics in the strong coupling and non-Markovian regime based on a reaction coordinate mapping.''  New Journal of Physics, 18(7), 073007.

\bibitem{Strong-Aurell-2018}
Aurell, Erik. ``Unified picture of strong-coupling stochastic thermodynamics and time reversals.'' Physical Review E 97.4 (2018): 042112.

\bibitem{Strong-Anders-2017}
Miller, Harry JD, and Janet Anders. ``Entropy production and time asymmetry in the presence of strong interactions.'' Physical Review E 95.6 (2017): 062123.

\bibitem{Gelin-2009}
Gelin, Maxim F., and Michael Thoss. ``Thermodynamics of a sub-ensemble of a canonical ensemble.'' Physical Review E 79.5 (2009): 051121.


\bibitem{Strong-Miguel-2020}
de Miguel, Rodrigo, and J. Miguel Rubí. ``Strong Coupling and Nonextensive Thermodynamics.'' Entropy 22.9 (2020): 975. 


\bibitem{Esposito-strong-coupling-2017}
Strasberg, Philipp, and Massimiliano Esposito. ``Stochastic thermodynamics in the strong coupling regime: An unambiguous approach based on coarse graining.'' Physical Review E 95.6 (2017): 062101.

\bibitem{Strasberg-2020-HMF}
Strasberg, Philipp, and Massimiliano Esposito. ``Measurability of nonequilibrium thermodynamics in terms of the Hamiltonian of mean force." Physical Review E 101.5 (2020): 050101.

\bibitem{Campisi-2009-HMF}
Campisi, Michele, Peter Talkner, and Peter Hänggi. ``Fluctuation theorem for arbitrary open quantum systems.'' Physical review letters 102.21 (2009): 210401.

\bibitem{Strong-Hsiang-20121}
Hsiang, Jen-Tsung, and Bei-Lok Hu. ``Zeroth law in quantum thermodynamics at strong coupling: In  equilibrium , not at equal temperature.'' Physical Review D 103.8 (2021): 085004.
%%%%%% quantum systems


\bibitem{Strong-Huang-20120}
Huang, Wei-Ming, and Wei-Min Zhang. ``Strong Coupling Quantum Thermodynamics with Renormalized Hamiltonian and Temperature.'' arXiv preprint arXiv:2010.01828 (2020).


\bibitem{Strong-Carrega-2016}
Carrega, Matteo, et al. ``Energy exchange in driven open quantum systems at strong coupling.'' Physical review letters 116.24 (2016): 240403.

\bibitem{Rivas-2020-strong-coupling}
Rivas, Angel. ``Strong coupling thermodynamics of open quantum systems.'' Physical Review Letters 124.16 (2020): 160601. 

\bibitem{Perarnau-Llobet-2018-strong-coupling}
Perarnau-Llobet, Martí, et al. ``Strong coupling corrections in quantum thermodynamics.'' Physical review letters 120.12 (2018): 120602.

\bibitem{Strasberg-2019-strong-coupling}
Strasberg, Philipp. ``Repeated interactions and quantum stochastic thermodynamics at strong coupling.'' Physical review letters 123.18 (2019): 180604.

\bibitem{Pucci-2013-poised}
Pucci, Lorenzo, Massimiliano Esposito, and Luca Peliti. ``Entropy production in quantum Brownian motion.'' Journal of Statistical Mechanics: Theory and Experiment 2013.04 (2013): P04005.

\bibitem{Esposito-2015-heat}
Esposito, Massimiliano, Maicol A. Ochoa, and Michael Galperin. ``Nature of heat in strongly coupled open quantum systems.'' Physical Review B 92.23 (2015): 235440.

\bibitem{Abe-2003-strong-coupling}
Abe, Sumiyoshi, and A. K. Rajagopal. ``Validity of the second law in nonextensive quantum thermodynamics.'' Physical review letters 91.12 (2003): 120601.


%\bibitem{Hanggi-Ingold-2008}
%Hänggi, Peter, Gert-Ludwig Ingold, and Peter Talkner. ``Finite quantum dissipation: the challenge of obtaining specific heat.'' New Journal of Physics 10.11 (2008): 115008.
%%%%%%%%%

\bibitem{Ptaszynski-2019}
Ptaszyński, Krzysztof, and Massimiliano Esposito. ``Entropy production in open systems: The predominant role of intraenvironment correlations.'' Physical Review Letters 123.20 (2019): 200603. 

\bibitem{ELB-2010}
Esposito, Massimiliano, Katja Lindenberg, and Christian Van den Broeck. ``Entropy production as correlation between system and reservoir.'' New Journal of Physics 12.1 (2010): 013013.

\bibitem{Miguel-Rubi-2020}
de Miguel, Rodrigo, and J. Miguel Rub\'i. ``Statistical Mechanics at Strong Coupling: A Bridge between Landsberg’s Energy Levels and Hill’s Nanothermodynamics.'' Nanomaterials 10.12 (2020): 2471.

\bibitem{Roux-solvent-1999}
Roux, Benoıt, and Thomas Simonson. ``Implicit solvent models.'' Biophysical chemistry 78.1-2 (1999): 1-20.

%%%%%%%%%%%%%%%%%%%%%%%%%%%%%%%%%%


\bibitem{Born-Fock-1928-adiabatic}
Born, M., and V. Fock. ``Proof of the adiabatic theorem.'' Z. Phys 51 (1928): 165-169.

\bibitem{Born-Oppenheimer-1927}
Max Born; J. Robert Oppenheimer (1927). "Zur Quantentheorie der Molekeln" [On the Quantum Theory of Molecules]. Annalen der Physik (in German). 389 (20): 457–484.

\bibitem{Gardiner-book}
Gardiner, Crispin W. {Handbook of Stochastic Methods,} {3rd ed.,} Berlin: Springer, {2004}.

\bibitem{Michaelis-Menten-1913}
Michaelis L \& Menten M (1913) Die kinetik der Invertinwirkung. Biochem Z 49, 333–369.

\bibitem{Pavliotis-Stuart-book-2008}
Pavliotis, Grigoris, and Andrew Stuart. ``Multiscale methods: averaging and homogenization.'' Springer Science \& Business Media, 2008.

\bibitem{Esposito-2010}
M. Esposito, K. Lindenberg, and C. Van den Broeck, Entropy
production as correlation between system and reservoir, New J.
Phys. 12, 013013 (2010).

\bibitem{Cover-Thomas}
Cover, Thomas M., and Joy A. Thomas. ``Elements of information theory''. John Wiley \& Sons, 2012. 

%%%%%%%%%%%%%%%%%%%%%%%%%%%%%%%%%%

%\bibitem{Martinez-Brownian-engine}
%Martínez, Ignacio A., et al. ``Brownian carnot engine.`` Nature physics 12.1 (2016): 67-70.

%\bibitem{Serreli-information-ratchet}
%Serreli, Viviana, et al. ``A molecular information ratchet.'' Nature 445.7127 (2007): 523-527.

%%%%%%%%%%%%%%%%%%%%%%%%%%%%%%%%%%

\bibitem{Kirkwood-1935}
Kirkwood, John G. ``Statistical mechanics of fluid mixtures.'' The Journal of chemical physics 3.5 (1935): 300-313.

\bibitem{covariant-Langevin-2020}
Ding, Mingnan, Zhanchun Tu, and Xiangjun Xing. ``Covariant formulation of nonlinear Langevin theory with multiplicative Gaussian white noises.'' Physical Review Research 2.3 (2020): 033381.

\bibitem{covariant-Thermodynamics-2021}
Ding, Mingnan  and Xiangjun Xing. ``Covariant Non-EQ Thermodynamics for Small Systems.'' 2021 arXiv:2105.14534. Submitted to Physical Review Research. 

\bibitem{Gibbs-book}
Gibbs, J. Willard. Elementary principles in statistical mechanics. Courier Corporation, 2014.

\bibitem{Esposito-2012}
Esposito, Massimiliano. "Stochastic thermodynamics under coarse graining." Physical Review E 85.4 (2012): 041125.

%\bibitem{Rushbrooke-1940}
%Rushbrooke, G.S. ``On the statistical mechanics of assemblies wose energy-levels depend on temperature.'' Trans. Faraday Soc. 1940, 36, 1055. 

%\bibitem{Landsberg-1954}
%Landsberg, P.T. ``Statitical Mechanics of Teperature-Dependent Energy Levels.'' Phys. Rev. 1954, 95, 643.

%\bibitem{Landsberg-1957}
%Elcock, E.W.; Landsberg, P.T. ``Temperature Dependent Energy Levels in Statistical Mechanics.'' Proc. Phys. Soc. Lond. Sect. B 1957, 70, 161. 


\end{thebibliography}
\end{document}